# A selective relaxation method for numerical solution of Schrödinger problems

Carlo Presilla
*Dipartimento di Fisica, Università di Roma "La Sapienza", and INFN, sezione di Roma, Piazzale A. Moro 2, Roma, Italy 00185*

Ubaldo Tambini
*Dipartimento di Fisica, Università di Ferrara, and INFN, sezione di Ferrara, Via Paradiso 12, Ferrara, Italy 44100*

We propose a numerical method for evaluating eigenvalues and eigenfunctions of Schrödinger operators with general confining potentials. The method is selective in the sense that only the eigenvalue closest to a chosen input energy is found through an absolutely-stable relaxation algorithm which has rate of convergence infinite. In the case of bistable potentials the method allows one to evaluate the fundamental energy splitting for a wide range of tunneling rates.

According to the von Neumann theory [1], in an ideal measurement of energy the state of a quantum system collapses instantaneously and completely into some eigenstate of the Hamiltonian. If one knows how to handle the collapse mechanism and how to select the final eigenstate, measurements of energy may be used for determining the spectrum of quantum systems. The restricted Feynman path-integral approach to quantum measurements [2] offers this possibility. During a continuous measurement of energy with known result $E$ (we consider the particular case of $E$ constant) the Feynman paths far from those compatible with the measurement result are damped proportionally to the accuracy of the measurement itself [3]. Choosing a damping of Gaussian type we obtain a Schrödinger problem with an effective Hamiltonian

$$H_{eff} = H - i\hbar\kappa(H-E)^2 \tag{1}$$

where the $\kappa$ gives the strength of the coupling to the measurement apparatus. In collaboration with R. Onofrio [4] we have recently discussed the dynamics of the wavefunction collapse induced by the effective Hamiltonian (1). Let us consider the case of $H$ with a nondegenerate discrete spectrum

$$H\psi_n(x) = E_n\psi_n(x) \tag{2}$$

where in units $\hbar = 2m = 1$

$$H = -\nabla_x^2 + V(x) \tag{3}$$

with $x \in I\!\!R^3$. The wavefunction of the measured system can be decomposed in terms of the eigenfunctions $\psi_n$ which are also eigenfunctions of $H_{eff}$. Due to the presence of the anti-Hermitian term in (1), during the measurement the initial wavefunction converges, up to a normalization factor, to the eigenfunction with energy closest to $E$ at a rate exponentially proportional to $\kappa$.

In principle, the numerical solution of the time-dependent Schrödinger equation with the effective Hamiltonian (1) represents a relaxation method for evaluating eigenvalues and eigenfunctions of $H$ close to the selecting energy $E$. In practice, we can speed up the relaxation by letting $\kappa \to \infty$. In this case we obtain the evolution equation

$$\frac{d}{dt}\psi(x,t;E) = -(H-E)^2\psi(x,t;E) \tag{4}$$

where the scaled time $t$ has dimensions $[E^{-2}]$. The function $\psi(x,t;E)$ is complex in general and we have emphasized its dependence on the selecting energy $E$. If we define the relaxed wavefunction and energy

$$\psi_{rel}(x;E) = \lim_{t\to\infty} \frac{\psi(x,t;E)}{||\psi(x,t;E)||} \tag{5}$$

$$E_{rel}(E) = \int \psi_{rel}(x;E)^* H \psi_{rel}(x;E)\, dx \tag{6}$$

they have the property that $\psi_{rel}(x;E) = \psi_n(x)$ and $E_{rel}(E) = E_n$ when $E \in \Gamma_n \equiv \,](E_n + E_{n-1})/2, (E_n + E_{n+1})/2[$ for $n \neq 0$. Relaxation to the ground state $n = 0$ is obtained through the weaker condition $E \in \Gamma_0 \equiv \,]-\infty, (E_0+E_1)/2[$.

It is worth noting the relevance of selectivity. We can evaluate whatever eigenstate of the spectrum just giving an estimate of it up to an error of the order of the local energy spacing. On the other hand, nonselective relaxation methods, like those based on the heat equation [5]

$$\frac{d}{dt}\psi = -H\psi \ , \tag{7}$$

converge only at the ground state (we suppose $H > 0$). Excited states can be obtained by using an initial wavefunction orthogonal to all the lower-energy states. However, only exactly orthogonal wavefunctions ensure relaxation to the desired eigenfunction. The errors introduced by finite-accuracy numerical orthogonalizations make the method unstable and not practical for determining high energy states.

Beside selectivity, another advantage characterizes a relaxation method based on Eq. (4). We can solve that equation through a finite-difference algorithm which is absolutely stable and allows us to evaluate the relaxed quantities (5-6) in one step. Let us explain the idea in the one dimensional case $x \in I\!R$. The domain of $x$ can be restricted to some interval $[x_{min}, x_{max}]$, depending on the selection energy $E$ and the confining potential $V(x)$, outside which the relaxed wavefunction vanishes within the computer accuracy. The interval is discretized by introducing the space lattice

$$x \to x_j = x_{min} + j\Delta x \qquad j = 0, 1, 2, \ldots, J+1 \tag{8}$$

where $(J+1)\Delta x = x_{max} - x_{min}$. If the time also is discretized according to

$$t \to t_m = m\Delta t \qquad m = 0, 1, 2, \ldots \tag{9}$$

Eq. (4) can be reduced to the following set of finite-difference equations

$$\frac{\psi_j^{m+1} - \psi_j^m}{\Delta t} = -(\gamma \psi_{j+2}^{m+1} + \beta_j^- \psi_{j+1}^{m+1} + \alpha_j \psi_j^{m+1} + \beta_j^+ \psi_{j-1}^{m+1} + \gamma \psi_{j-2}^{m+1}) \tag{10}$$

where

$$\gamma = \frac{1}{\Delta x^4} \tag{11}$$

$$\beta_j^{\pm} = -\frac{4}{\Delta x^4} + \frac{2}{\Delta x^2}(E - V_j) \pm \frac{1}{\Delta x} V_j' \tag{12}$$

$$\alpha_j = \frac{6}{\Delta x^4} - \frac{4}{\Delta x^2}(E - V_j) + (E - V_j)^2 - V_j'' \tag{13}$$

and $V_j$, $V_j'$ and $V_j''$ are the values of the potential and its first two space-derivatives at $x_j$. Due to the boundary conditions we can rewrite (10) in a compact form suitable for numerical solution

$$\mathcal{R}_{ij} \psi_j^{m+1} = \psi_i^m \tag{14}$$

where the matrix $\mathcal{R}$ is pentadiagonal with nonvanishing elements $\mathcal{R}_{ii} = 1 + \Delta t\ \alpha_i$, $\mathcal{R}_{ii\pm1} = \Delta t\ \beta_i^{\mp}$ and $\mathcal{R}_{ii\pm2} = \Delta t\ \gamma$. Starting with a known $\psi_j^0$ the system (14) is efficiently solved with standard decomposition and back-substitution method [5] in a number of operations per time step proportional to $J$.

Following the von Neumann stability analysis [5], the eigenmodes $\psi_j^m(k) = \xi^m e^{ikj\Delta x}$ substituted back into (14) give

$$\left|\frac{\xi^{m+1}}{\xi^m}\right|^2 = \frac{1}{|1 + (a + ib)\Delta t|^2} \tag{15}$$

where

$$a = (V_j - E)^2 + \frac{8}{\Delta x^2}(V_j - E)\sin^2\left(\frac{k\Delta x}{2}\right) + \frac{16}{\Delta x^4}\sin^2\left(\frac{k\Delta x}{2}\right) - \frac{4}{\Delta x^4}\sin^2(k\Delta x) - V_j'' \tag{16}$$

$$b = -\frac{2}{\Delta x} V_j' \sin(k\Delta x)\ . \tag{17}$$

Stability is obtained when the growing ratio (15) is smaller than unity, i.e. when

$$\Delta t \geq -\frac{2a}{a^2 + b^2}\ . \tag{18}$$

This means that we can choose $\Delta t$ very large and obtain convergence to the relaxed quantities (5-6) in a single iteration of Eq. (14). Differently stated the rate of convergence for the recursive equation $\psi^{m+1} = \mathcal{R}^{-1}\psi^m$ is infinite. This can be seen directly by the formula for the rate of convergence [6], $-\ln S(\mathcal{R}^{-1})$, where $S(\ldots)$ means spectral radius. In the limit $\Delta t \to \infty$ the eigenvalues of $\mathcal{R}^{-1}$ and its spectral radius vanish and the rate of convergence diverges. Finite computer accuracy imposes a limitation on the value of $\Delta t$. The relaxed wavefunction is obtained after normalization of the vanishing $\psi(x, t; E)$ and $\Delta t$ cannot be so large that $\psi_j^1$ yields underflow. The limitation, however, is not crucial and full relaxation can be usually obtained with very few iterations (we never use more than ten iterations) even without making an optimal choice for $\Delta t$ (maximum value allowed).

Efficiency and precision obtainable from the selective relaxation method [7] have been checked in various cases. These include exactly solvable problems as well as problems where results obtained with different numerical procedures are available for comparison [8,9]. Here we report only on a comparison with the exact results of the Morse potential and on the possibility to evaluate with great accuracy the fundamental energy splitting of double-well potentials.

The eigenvalue problem with the Morse potential

$$V(x) = e^{-2\mu x} - 2e^{-\mu x} \tag{19}$$

has well known analytical solutions [10]. In Fig.s 1 and 2 we compare the exact eigenvalues and eigenfunctions with the corresponding energies and wavefunctions obtained with the selective relaxation method for different values of the lattice step $\Delta x$. We chose $\mu = 0.2$ which corresponds to have five bound states $n = 0, \ldots, 4$. The convergence to the $n$th bound state is absolutely insensitive to the choice of the initial wave function as well as of the selecting energy $E$ in the interval $\Gamma_n$.

According to the discretization scheme used in (10) the algorithm is first-order accurate in the lattice step $\Delta x$. More explicitly, for the eigenvalues we observe a systematic error

$$E_{rel} - E_n = \epsilon_n \Delta x^2 \tag{20}$$

with $\epsilon_n$ increasing with $n$. Since the computation time is proportional to the number of steps in the space lattice, we conclude that the error in the evaluation of the eigenvalues decreases quadratically with the computation time. In order to fix the ideas, the computation time for $J = 10^4$, which is a typical figure in evaluating $E_4$ with a 0.01 % error, is about 2 s per time iteration in a 25 MHz 486 PC.

The increase of $\epsilon_n$ with $n$ may become a problem when evaluating high energy eigenstates. Indeed, the dimension $J$ of the space lattice necessary for controlling the error (20) through very small $\Delta x$ may exceed the computer capacity. The problem is overcome by resorting to a higher-order approximation in the discretization of the operator $(H - E)^2$. If we substitute the right hand side of Eq. (10) with a $k$th-order accurate expression the error (20) gets proportional to $\Delta x^{k+1}$ and we have higher precision for a given lattice dimension $J$. In this case, however, the matrix $\mathcal{R}$ has $2(k + 1) + 1$ nonvanishing diagonals (for $x \in I\!\!R$) and the computation time for the same $J$ increases [5]. A quantitative comparison between the minimal-accuracy algorithm presented here and higher-accuracy ones, also in the cases $x \in I\!\!R^2$ and $x \in I\!\!R^3$ is, deferred elsewhere.

Even in its minimal-accuracy version the selective relaxation method allows us to make a relevant advance in the evaluation of the fundamental energy splitting $T = E_1 - E_0$ of a double-well potential. It is well known that this problem gets rapidly unapproachable with standard numerical methods when the tunneling rate between the two wells decreases [11]. A different situation arises for the fundamental energy splitting $T_{rel}$ obtained as difference of the lowest two relaxed eigenvalues

$$T_{rel} = T + (\epsilon_1 - \epsilon_0)\Delta x^2 \ . \tag{21}$$

$T_{rel}$ shows a systematic error due to the finite lattice, $(\epsilon_1 - \epsilon_0)\Delta x^2$, which is the difference of two close numbers and vanishes for $T$ vanishing. In addition, the condition $E_0 \simeq E_1$ causes no trouble in selecting the two eigenvalues since the corresponding eigenfunctions $\psi_0$ and $\psi_1$ have different parity. Initial wavefunctions $\psi(x, 0; E)$ with the same selecting energy, e.g. chosen as the WKB approximation to the ground state, but different parity automatically relax toward the eigenfunctions with the corresponding parity.

An example of the discussed behavior is shown in Fig. 3. We have considered the bistable potential

$$V(x) = -\lambda x^2 + x^4 \tag{22}$$

for $\lambda = 15$. The values of $T_{rel}$ obtained for different values of the space step $\Delta x$ follows accurately law (21). A linear fit gives $T = 1.9496 \times 10^{-10}$ and $\epsilon_1 - \epsilon_0 = 2.6404 \times 10^{-8}$. This last figure should be compared with the single level accuracy $\epsilon_0 \simeq \epsilon_1 \simeq 7.22$ which is eight orders of magnitude greater.

Comparison of the splitting values obtained through selective relaxation with those obtained through other techniques establishes a clear superiority of our method. Recently a substantial advance in evaluating the fundamental energy splitting of double-well potentials was realized with the technique of supersymmetric quantum mechanics [11]. Using as input the ground-state wavefunction $\psi_0(x)$, obtained, for instance, with standard Runge-Kutta integration, the supersymmetric approach allows one to evaluate the splitting through a logarithmic perturbation series which converges rapidly in the limit of small tunneling rate. In Table I we compare the fundamental energy splitting of the double-well potential (22) evaluated with the Runge-Kutta method, the supersymmetric series at third order and the selective relaxation for different values of $\lambda$. For $\lambda$ small the potential is weakly bistable and the Runge-Kutta method is reliable. The supersymmetric result shows weak convergence in this limit. For $\lambda$ large the Runge-Kutta method is unreliable for evaluating the splitting. However, this method is still good for evaluating the ground state used in the supersymmetric series which shows good convergence. The selective relaxation method gives the right result in the full range of $\lambda$ values with a minimal amount of computation time. Notice that, in addition to the cases considered in Ref. [11], we are able to evaluate the splitting for $\lambda = 15$, i.e. in a region where $T$ is close to the maximal accuracy available in our computer (double precision).

Some final comments are in order. We have considered only Hamiltonians with nondegenerate spectra. The selective relaxation method, however, applies also in presence of degeneracy which may occur in two- or three-dimensional cases. Initial wavefunctions which are mutually orthogonal converge to the corresponding degenerate eigenfunctions. In fact, this property was used for selecting the nearly-degenerate fundamental levels of the double-well potential (22).

Modified boundary conditions, e.g. asymptotic known values at the points $x_{min}$ and $x_{max}$, allow one to evaluate eigenfunctions in the continuous spectrum. In particular, extended resonance states can be found through the so called quantum transmitting boundary method [12]. Periodic boundary conditions allow one to calculate eigenvalues and eigenfunctions of periodic potentials.

The main characteristics of the proposed relaxation method, namely selectivity and absolute stability, can be extended to different classes of eigenvalue problems such as those involving Fokker-Planck and Dirac operators.

## ACKNOWLEDGMENTS

We thank G. Jona Lasinio and R. Onofrio for valuable discussions and F. Marchesoni and R. Onofrio for a critical reading of the manuscript. This work was supported in part by INFN Iniziativa Specifica RM6.

TABLE I. Fundamental energy splitting of the double-well potential $V(x) = -\lambda x^2 + x^4$ obtained in Ref. [11] with Runge-Kutta integration, $T_{RK}$, and supersymmetric perturbation series at third order, $T_{SS3}$, compared with the selective relaxation result, $T_{rel}$, at lattice step $\Delta x = 10^{-3}$. Notice that numerical Runge-Kutta calculations are unreliable for $\lambda \geq 10$ while supersymmetric result gets inaccurate for $\lambda$ small.

| $\lambda$ | $T_{RK}$ | $T_{SS3}$ | $T_{rel}$ |
|---|---|---|---|
| 0.5 | 2.464 | 2.451 | 2.4637 |
| 1 | 2.177 | 2.168 | 2.1769 |
| 2 | 1.575 | 1.573 | 1.5752 |
| 3 | $9.712 \times 10^{-1}$ | $9.712 \times 10^{-1}$ | $9.7115 \times 10^{-1}$ |
| 4 | $4.624 \times 10^{-1}$ | $4.624 \times 10^{-1}$ | $4.6242 \times 10^{-1}$ |
| 5 | $1.595 \times 10^{-1}$ | $1.595 \times 10^{-1}$ | $1.5947 \times 10^{-1}$ |
| 6 | $4.14 \times 10^{-2}$ | $4.14 \times 10^{-2}$ | $4.1398 \times 10^{-2}$ |
| 7 | $8.65 \times 10^{-3}$ | $8.65 \times 10^{-3}$ | $8.6531 \times 10^{-3}$ |
| 8 | $1.52 \times 10^{-3}$ | $1.52 \times 10^{-3}$ | $1.5164 \times 10^{-3}$ |
| 9 | $2.28 \times 10^{-4}$ | $2.28 \times 10^{-4}$ | $2.2792 \times 10^{-4}$ |
| 10 | $2.86 \times 10^{-5}$ | $2.98 \times 10^{-5}$ | $2.9821 \times 10^{-5}$ |
| 11 | | $3.43 \times 10^{-6}$ | $3.4338 \times 10^{-6}$ |
| 12 | | $3.51 \times 10^{-7}$ | $3.5093 \times 10^{-7}$ |
| 15 | | $1.95 \times 10^{-10}$ [a] | $1.9499 \times 10^{-10}$ |

[a] first-order supersymmetric result using as input the ground state obtained with selective relaxation

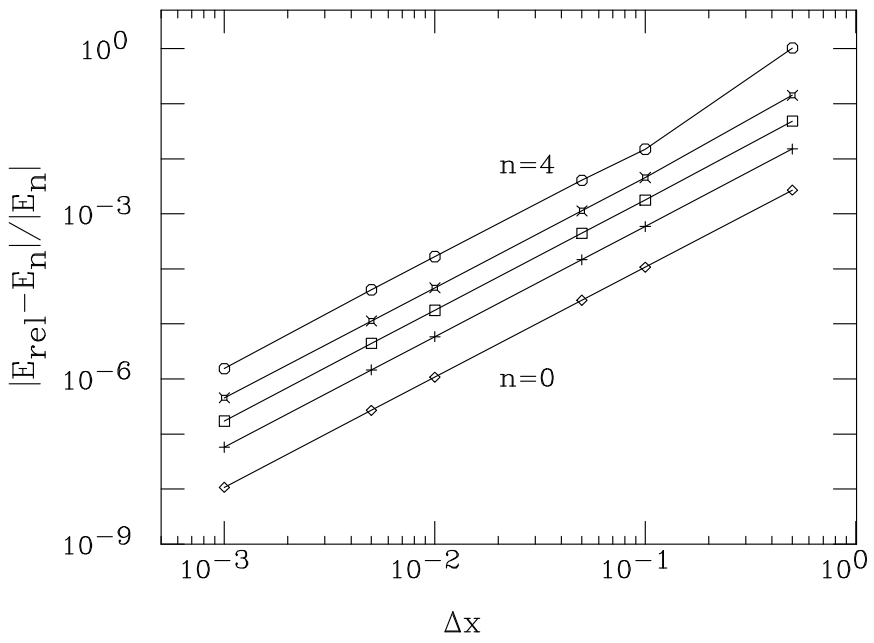

FIG. 1. Comparison of the exact eigenvalues $E_n$ of the Morse potential with the relaxed eigenvalues $E_{rel}$ for different values of the lattice step $\Delta x$. The potential is $V(x) = e^{-2\mu x} - 2e^{-\mu x}$ with $\mu = 0.2$ and has five bound states $n = 0, \ldots, 4$.

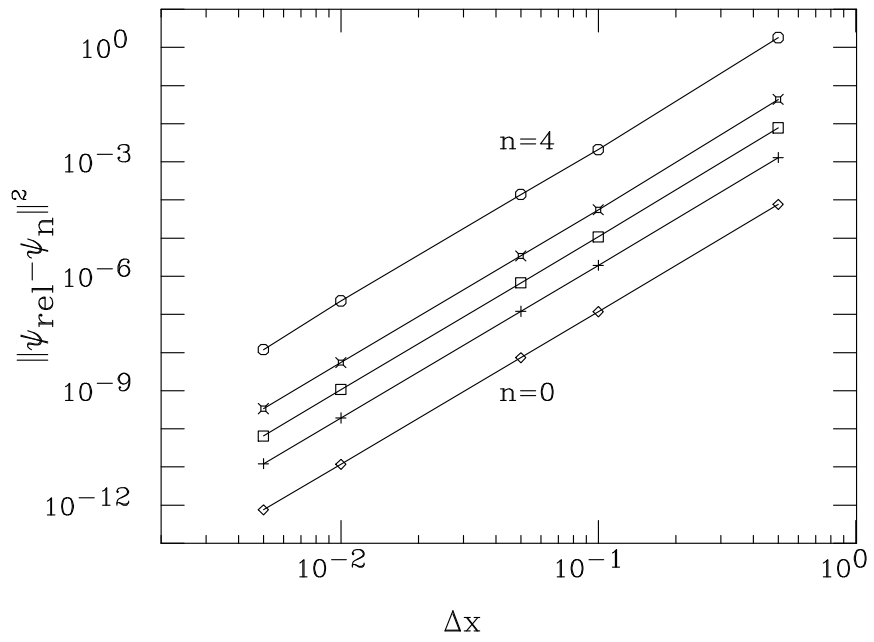

FIG. 2. Comparison of the exact eigenfunctions $\psi_n$ of the potential of Fig. 1 with the relaxed eigenfunctions $\psi_{rel}$ for different values of the lattice step $\Delta x$.

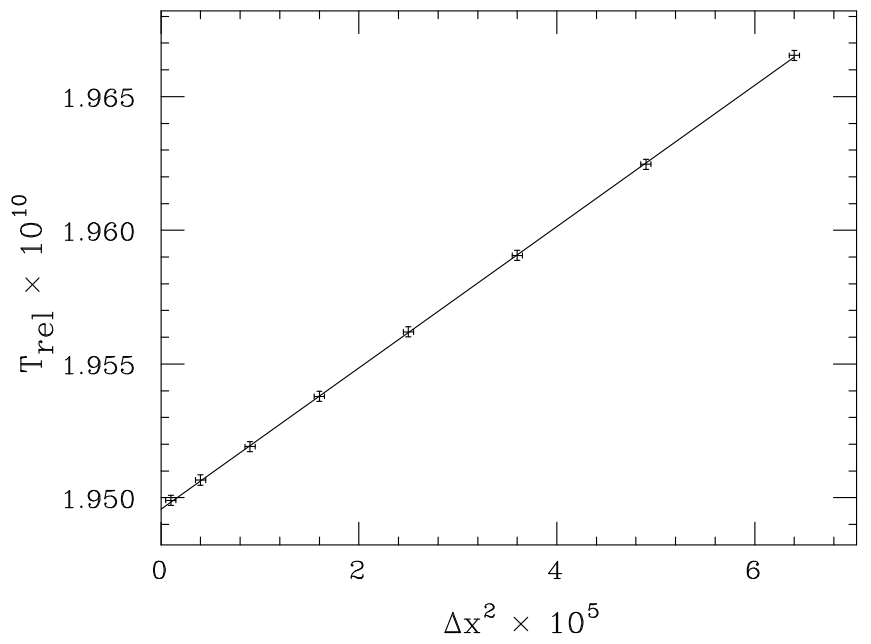

FIG. 3. Fundamental energy splitting $T_{rel}$ obtained as difference of the lowest two relaxed eigenvalues of the potential $V(x) = -15x^2 + x^4$ for different values of the lattice step $\Delta x$. A linear fit (solid line) gives $T_{rel} = 1.9496 \times 10^{-10} + 2.6404 \times 10^{-8} \Delta x^2$.